\documentstyle[seceq,epsf]{ptptex}

\title{Non-Abelian Stokes Theorem for Loop Variables\\
Associated with Nontrivial Loops}

\author{Minoru {\sc Hirayama}, Mika {\sc Kanno}, 
Masataka {\sc Ueno} and Hitoshi {\sc Yamakoshi}$^*$}

\inst{Department of Physics, Toyama University, Toyama 930-8555, Japan
\\
$^*$Toyama National College of Technology, Toyama 939-8630, Japan
}

\recdate{\today}

\abst
{The non-abelian Stokes theorem for loop variables
associated with nontrivial loops (knots and links) is derived.
It is shown that a loop variable is in general different
from unity even if the field strength vanishes everywhere
on the surface surrounded by the loop.}

\begin{document}

\maketitle

\section{Introduction}

As is well known, the Stokes theorem is remarkable for 
its generality : the formula

\begin{equation}
\int_{\partial M}\omega=\int_Md\omega
\end{equation}
is valid for any $(d+1)$-dimensional oriented manifold $M$ with the 
boundary $\partial M$, where $\omega$ is a $d$-form on $M$.
It should be noted that $\omega$ and $M$ in $(1\cdot 1)$ can be 
replaced with a $p$-form and a $(p+1)$-chain with $p\leq d$, respectively.
Its usefulness manifests in the following formula of electromagnetism :

\begin{equation}
\int_{\partial\sigma}a_\mu(x)dx^\mu
=\frac{1}{2}\int\hspace{-2mm}\int_\sigma f_{\mu\nu}(x)dx^\mu\wedge dx^\nu,
\end{equation}

\begin{equation}
f_{\mu\nu}(x)=\partial_\mu a_\nu(x)-\partial_\nu a_\mu(x),
\end{equation}
where $\sigma,a_\mu(x)$ and $f_{\mu\nu}(x)$ are a two-dimensional
oriented surface in the spacetime, the electromagnetic potential
and the electromagnetic field strength, respectively.\\
On the other hand, if we consider the non-Abelian gauge potential

\begin{equation}
A_\mu(x)=A^a_\mu(x)T^a,
\hspace{5mm}T^a:\mbox{\rm generator of the gauge group},
\end{equation}
and the field strength

\begin{equation}
F_{\mu\nu}(x)=\partial_\mu A_\nu(x)-\partial_\nu A_\mu(x)
-ig[A_\mu(x),A_\nu(x)],
\end{equation}
with $g$ being the gauge coupling constant, we have 
$\int_{\partial\sigma}A_\mu(x)dx^\mu
=\frac{1}{2}\int\hspace{-2mm}\int_\sigma
\bigl\{\partial_\mu A_\nu(x)-\partial_\nu A_\mu(x) \bigl\}
dx^\mu\wedge dx^\nu
\neq\frac{1}{2}\int\hspace{-2mm}\int_\sigma F_{\mu\nu}(x)dx^\mu\wedge dx^\nu.$
We see that a line integral of the gauge potential cannot be converted to 
a surface integral of the field strength  in non-Aberian gauge theory.
 
An important variable of the non-Abelian gauge field theory is the loop 
variable $(\gamma)$ defined by\cite{rf:1}\tocite{rf:5}

\begin{equation}
\left.\begin{array}{l}
(\gamma)=P_\kappa e^{ig\int_0^1d\kappa A_\mu(x(\kappa))
\frac{dx^\mu(\kappa)}{d\kappa}},\\
\\
P_\kappa :\hspace{3mm}\kappa\mbox{\rm -ordering},\\
\end{array}\right.
\end{equation}
where the loop $\gamma$ is parametrized by the parameter $\kappa$ as 
$\gamma=\{x(\kappa)|0\leq\kappa\leq1\}$.
Under the gauge transformation 

\begin{equation}
A_\mu(x)\rightarrow {A^{'}}_\mu(x)=h(x)A_\mu(x)h^{-1}(x)
+\frac{i}{g}h(x)\partial_\mu h^{-1}(x),
\end{equation}
the loop variable $(\gamma)$ transforms as 
 
\begin{equation}
(\gamma)\rightarrow h(x(0))(\gamma)h^{-1}(x(0)),
\end{equation}
where the point $x(0)$ is the starting point of $\gamma$, which should
coincide with the end point $x(1)$. When the loop $\gamma$ is 
trivial,i.e., unknotted and unlinked, the loop variable $(\gamma)$ is 
equal to the following quantity :

$$[S]\equiv P_te^{ig\int_0^1dt\int_0^1ds{\mathcal F_{\mu\nu}}(x(s,t))
\frac{\partial x^\mu(s,t)}{\partial s}
\frac{\partial x^\nu(s,t)}{\partial t}},$$
\begin{equation}
{\mathcal F_{\mu\nu}}(x)=w(x)F_{\mu\nu}(x)w^{-1}(x),
\end{equation}
$$P_t\hspace{3mm}:\hspace{5mm}t\mbox{\rm -ordering},$$
where $w(x)$ is a unitary matrix depending on a path from 
$x(0,0)$ to $x(s,t)$ and the boundary $\partial S$
of the simply connected surface $S=\{x(s,t)|(s,t)\in\Sigma\}$,
$\Sigma=\{(s,t)|0\leq s,t\leq1\}$, is assumed to be equal to 
the loop $\gamma$. The $x(0)=x(1)$ in (1$\cdot$6) sould 
coincide with the $x(0,0)$ in (1$\cdot$9). The equality

\begin{equation}
(\gamma)=[S],\hspace{5mm}\gamma=\partial S
\end{equation}
is called the non-Abelian Stokes theorem(NAST).\cite{rf:6}\tocite{rf:15}
For a given loop $\gamma$,
there exist many surfaces satisfying $\partial S=\gamma$, which are 
continuously deformable to each other. 
It was shown that the Bianchi identity 

\begin{equation}
\left.\begin{array}{l}
[D_\rho,F_{\mu\nu}]+[D_\mu,F_{\nu\rho}]+[D_\nu,F_{\rho\mu}]=0,\\
\\
D_\rho=\partial_\rho-igA_\rho,\\
\end{array}\right.
\end{equation}
guarantees the invariance of $[S]$ under continuous deformations of 
$S$.\cite{rf:15} In other words, for the variation

\begin{equation}
x(s,t)\rightarrow x(s,t)+\delta x(s,t),
\end{equation}
with the property 

\begin{equation}
\delta x(s,t)=0,\hspace{2mm}(s,t)\in\partial\Sigma,
\end{equation}
we have 

\begin{equation}
\delta[S]=0.
\end{equation}
In the above, we have stated the NAST (1$\cdot$10)  
under the assumption that the loop $\gamma$ is trivial and surrounds 
the simply connected surface $S$. If the loop $\gamma$ is nontrivial
and the surface $S$ is not simply connected, the parameter space $\Sigma$
must be chosen more complicated than the above 
$\{(s,t)|0\leq s,t\leq1\}$.
The purpose of the present paper is to explore how the NAST should be 
modified when the loop $\gamma$ is a knot or a link.
We shall find that rather simple applications of the knot 
theory\cite{rf:16,rf:17} leads us to the NAST in such cases.
We shall also find that the loop variable $(\gamma)$ can be different 
from unity even if the field strength $F_{\mu\nu}(x)$ vanishes at 
every point on the surface $S$.

This paper is organized as follows. In \S2, we obtain the NAST for the case 
that the loop is a trefoil knot. After obtaining the NAST for a Hopf link
in \S3, we discuss the NAST for an arbitrary loop in \S4. The final section,
\S5, is devoted to summary.

\section{NAST for a trefoil knot}

Before considering the general case, we discuss in the present and the next 
sections loop variables associated with some simple but nontrivial loops.
Assuming that the loops$\subset${\bf\itshape R}$^3\hspace{-1mm}
\subset$spacetime, we begin with the case of a trefoil knot. 

\subsection{Surfaces surrounded by a trefoil knot}

The first task to be done is to find an oriented surface whose boundary is 
ambient isotopic, i.e., continuously deformable in {\bf\itshape R}${}^3$,
to a trefoil knot shown in Fig.1. There exists a standerd method called
the Seifert algorithm to construct surfaces with desired properties.

\begin{figure}
    \epsfxsize=4cm
    \centerline{\epsfbox{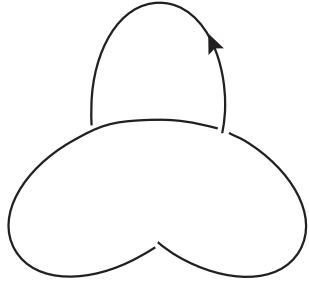}}
\caption{A trefoil knot.}
\label{Fig.1}
\end{figure}

It is not difficult to recognize that the surfaces $S$, $S'$ and $S''$
in Fig.2 are three such examples. 

\begin{figure}
    \epsfxsize=14cm
    \centerline{\epsfbox{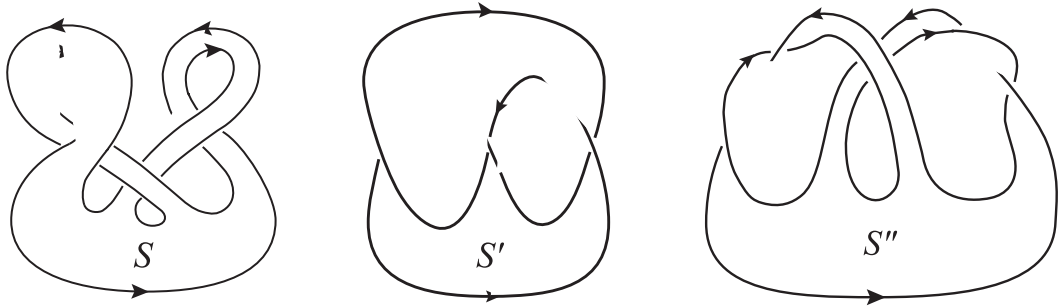}}
\caption{Seifert surfaces for loops ambient isotopic to a trefoil knot.}
\label{Fig.2}
\end{figure}

The surface $S$ is said to be of 
the Seifert standard form. It is clear that the 
surface $S$ is homeomorphic to the surface $\Sigma$ of Fig.3 :
there exist a continuous bijection $x:\Sigma\rightarrow S$
and $x^{-1}$ : $S\rightarrow\Sigma$ is also continuous.
As is explained in \S4, a surface whose boundary coincides with a given 
link is called the Seifert surface of the link. 
surface $S$ is homeomorphic to the surface $\Sigma$ of Fig.3 :
there exist a continuous bijection $x:\Sigma\rightarrow S$
and $x^{-1}$ : $S\rightarrow\Sigma$ is also continuous.
As is explained in \S4, a surface whose boundary coincides with a given 
link is called the Seifert surface of the link. 

\begin{figure}
    \epsfxsize=5cm
    \centerline{\epsfbox{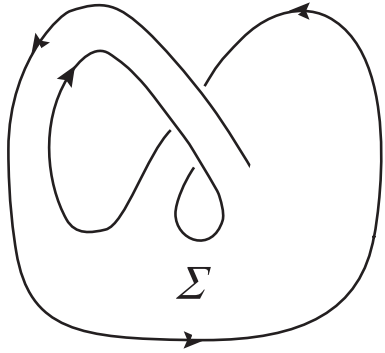}}
\caption{The parameter space for a Seifert surface of a trefoil knot.}
\label{Fig.3}
\end{figure}

We are allowed to regard 
the surface $S$ an oriented surface in the spacetime satisfying 
$\partial S=\gamma$ and assume $S=\{x(s,t)|(s,t)\in\Sigma\}$.
Here the surface $\Sigma$ plays the role of the parameter space which was 
necessary in the description of the NAST in \S1. Our procedure can be stated 
as follows. We first choose the parameter space $\Sigma$ to be of the 
allowed simplest structure. The surface $S$ embedded in the spacetime is 
given as $S=x(\Sigma)$, where the mapping $x$ may cause some twists 
and linkings of bands.

\subsection{Decomposition of $\Sigma$ into simply connected surfaces}

Although there are many ways to decompose the surface $\Sigma$ into 
some simply connected surfaces, we adopt the manner shown in Fig.4.

\begin{figure}
    \epsfxsize=10cm
    \centerline{\epsfbox{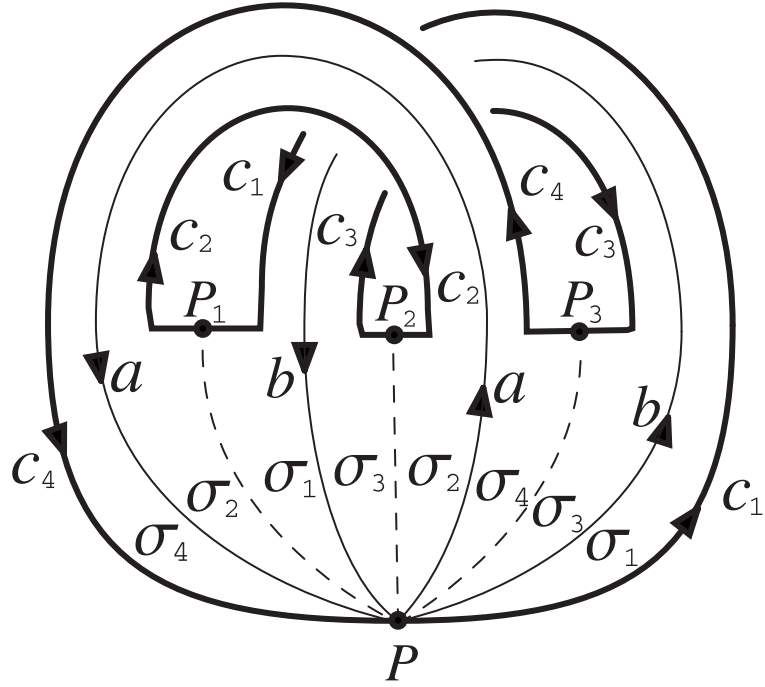}}
\caption{A decomposition of the parameter 
          space into simply connected surfaces.}
\label{Fig.4}
\end{figure}

The surfaces $\sigma_i,\hspace{1mm}i=1,2,3,4$, in Fig.4 satisfying 
$\Sigma=\bigcup_{i=1}^4\sigma_i$
correspond to the surfaces $S_i,i=1,2,3,4$, in the spacetime, respectively :

\begin{equation}
S=\bigcup_{i=1}^4S_i,\hspace{2mm}S_i=x(\sigma_i).
\end{equation}
Similarly the portions $c_i$, $i=1,2,3,4,$ of the boundary of $\Sigma$ correspond
to those of the trefoil knot $\gamma$ in the spacetime :

\begin{equation}
\gamma=\gamma_4\circ\gamma_3\circ\gamma_2\circ\gamma_1,\hspace{5mm} 
\gamma=x(c_i).
\end{equation}
The points $P_{i-1}$ and $P_i$ denote the starting and the end points of 
$c_i$, $i=1,2,3,4,$ $P_0\equiv P_4\equiv P$,respectively. 
The curves $a$ and $b$ are two independent elements of the 
first homology group of $\Sigma$.We have thus decomposed the surface 
$\Sigma$ into four simply connected surfaces 
$\sigma_i,\hspace{2mm}i=1,2,3,4,$ with the help of $a,b,d_1,d_2$
and $d_3$, where $d_i$ is a curve starting at $P_i$ and ending at $P$.

\subsection{Derivation of NAST}

The surface $\sigma_1$ is surrounded by the boundary 
$\overline{b}\circ d_1\circ c_1$, where $\overline{b}$ is $b$ with the 
orientation reversed. Since the surface $\sigma_1$ is simply connected,
we can apply the NAST of \S1 with $S=S_1=x(\sigma_1)$ and 
$\gamma=\partial S_1=\overline{B}\circ D_1\circ\gamma_1=x(\overline{b}
\circ d_1\circ c_1)$, where $\overline{B}$ and $D_i$ are defined by

\begin{equation}
\overline{B}=x(\overline{b}),\hspace{5mm}D_i=x(d_i).
\end{equation}
We then have 
 
\begin{equation}
[S_1]=(\overline{B})(D_1)(\gamma_1),
\end{equation}
where $[*]$ and $(*)$ are defined in analogous manners to $(1\cdot9)$
and $(1\cdot6)$, respectively. From  $(2\cdot4)$, we obtain

\begin{equation}
(\gamma_1)=(D_1)^{-1}(B)[S_1],
\end{equation}
where we have made use of the relation $(\overline{B})=(B)^{-1}$.
Similarly we have

\begin{equation}
\left.\begin{array}{l}
(A)(D_2)(\gamma_2)(\overline{D_1})=[S_2],\\
\\
(B)(D_3)(\gamma_3)(\overline{D_2})=[S_3],\\
\\
(\overline{A})(\gamma_4)(\overline{D_3})=[S_4],\\
\end{array}\right.
\end{equation}
yielding 

\begin{equation}
\left.\begin{array}{l}
(\gamma_2)=(D_2)^{-1}(A)^{-1}[S_2](D_1),\\
\\
(\gamma_3)=(D_3)^{-1}(B)^{-1}[S_3](D_2),\\
\\
(\gamma_4)=(A)[S_4](D_3),\\
\end{array}\right.
\end{equation}
with 

\begin{equation}
A=x(a).
\end{equation}
Recalling that the loop variable $(\gamma)$ is given by 

\begin{equation}
(\gamma)=(\gamma_4)(\gamma_3)(\gamma_2)(\gamma_1),
\end{equation}
and that the trivial loops $a$ and $b$ surround simply connected 
domains $\sigma_a$ and $\sigma_b$, respectively 
$(\partial\sigma_a=a$, $\partial\sigma_b=b)$, we are led to

\begin{equation}
\left.\begin{array}{l}
(\gamma)=[S_a][S_4][S_b]^{-1}[S_3][S_a]^{-1}[S_2][S_b][S_1],\\
\\
\hspace{6mm}\equiv[\hspace{-0.7mm}[S]\hspace{-0.7mm}]\\
\end{array}\right.
\end{equation}
where $S_a$ and $S_b$ are given by

\begin{equation}
S_a=x(\sigma_a),\hspace{2mm}S_b=x(\sigma_b).
\end{equation}
The l.h.s. of (2$\cdot$10) concerns a contour integral of the 
non-Abelian gauge potential $A_\mu(x)$, while the r.h.s of (2$\cdot$10) 
with surface integrals of field strength.
Eq. (2$\cdot$10) should be regarded as the NAST in the case that the loop 
$\gamma$ is a trefoil knot.

Some comments are in order.\\
(a) It is possible to think of a surface $\widetilde{S}$ which satisfies
$\partial\widetilde{S}=\gamma$ and is oriented but 
cannot be continuously deformed to
the above considered $S$. As will be discussed in \S4, 
it can be shown that $[\hspace{-0.7mm}[S]\hspace{-0.7mm}]$
equals $[\hspace{-0.7mm}[\widetilde{S}]\hspace{-0.7mm}]$.\\
(b) Although the r.h.s. of (2$\cdot$10) may seem to depend on the choice of
closed curves $a$ and $b$ on $\Sigma$, it is not the case : the r.h.s. of 
(2$\cdot$10) does not vary under small 
deformations of $a$ and $b$. This fact can be understood through the 
observations

\begin{equation}
\left.\begin{array}{l}
\delta_a\bigl([S_a]^{-1}[S_2]\bigl)
=\delta_a\bigl((D_2)(\gamma_2)(D_1)^{-1}\bigl)=0,\\
\\
\delta_b\bigl([S_b][S_1]\bigl)=\delta_b\bigl((D_1)(\gamma_1)\bigl)=0,
\end{array}\right.
\end{equation}
etc., where $\delta_a$ and $\delta_b$ imply small deformations of 
$a$ and $b$, respectively.\\ 
(c) Eq. (2$\cdot$10) can be rewritten as follows :

\begin{equation}
\left.\begin{array}{l}
(\gamma)={}^{g_4}[S_4]\hspace{2mm}{}^{g_3}[S_3]\hspace{2mm}{}^{g_2}[S_2]
\hspace{2mm}{}^{g_1}[S_1]\hspace{2mm}[\xi]\\
\\
\hspace{6mm}=[\xi]\hspace{2mm}{}^{h_4}[S_4]\hspace{2mm}{}^{h_3}[S_3]
\hspace{2mm}{}^{h_2}[S_2]\hspace{2mm}{}^{h_1}[S_1],\\
\end{array}\right.
\end{equation}
with 

\begin{equation}
[\xi]=[S_a][S_b]^{-1}[S_a]^{-1}[S_b],
\end{equation}

\begin{equation}
\left.\begin{array}{l}
{}^{g_i}[S_i]=g_i[S]g_i^{-1},\hspace{3mm}{}^{h_i}[S_i]=h_i[S_i]h_i^{-1},\\
\\
g_1=[\xi],\hspace{3mm}g_2=[S_a][S_b]^{-1}[S_a]^{-1},\\
\\
g_3=[S_a][S_b]^{-1},\hspace{3mm}g_4=[S_a],\\
\\
h_1=1,\hspace{3mm}h_2=[S_b]^{-1},\hspace{3mm}h_3=[S_b]^{-1}[S_a],\\
\\
h_4=[S_b]^{-1}[S_a][S_b].\\
\end{array}\right.
\end{equation}
(d)The parameter space of the type of Fig.3 can be used for loops other than 
the trefoil knot. For example, the figure eight knot shown in Fig.5(a) has 
the Seifert surface of Fig.5(b), which is homeomorphic to the $\Sigma$ 
of Fig.3.

\begin{figure}
    \epsfxsize=10cm
    \centerline{\epsfbox{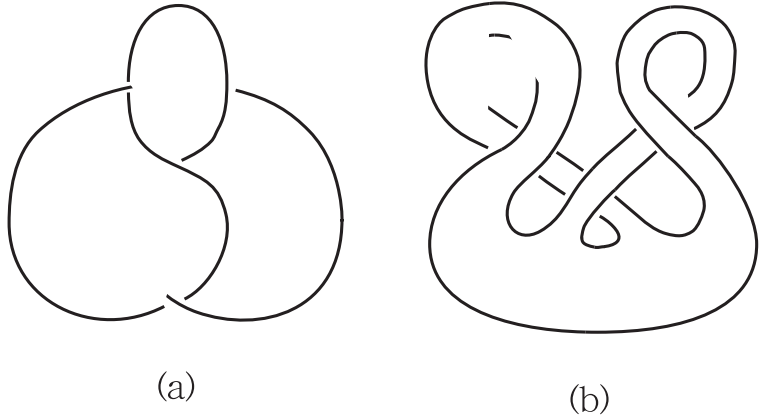}}
\caption{(a)A figure eight knot, (b)A Seifert surfaces of a 
          loop ambient isotopic to a figure eight knot.}
\label{Fig.5}
\end{figure}

(e) In the Abelian case, Eq. $(2\cdot10)$ reduces to 
$(\gamma)=[S_4][S_3][S_2][S_1]=[S]$.

\subsection{Example}

We consider the case that the loop $\gamma$ is the boundary of the surface
$S'''$ shown in Fig.6, which is a deformed version of $S''$ of Fig.2.

\begin{figure}
    \epsfxsize=11cm
    \centerline{\epsfbox{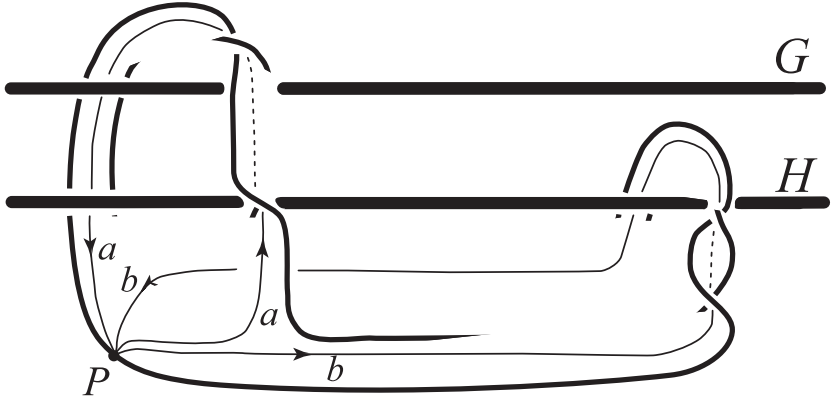}}
\caption{A deformed version of $S''$ of Fig.2.}
\label{Fig.6}
\end{figure}

We assume that $F_{\mu\nu}(x)$ does not vanish only in the neighbourhoods
of the lines $G$ and $H$. Especially, we assume

\begin{equation}
F_{\mu\nu}(x)=0,\hspace{3mm}x\in S'''.
\end{equation} 
Then we have in general

\begin{equation}
[S_a]\neq1,\hspace{3mm}[S_b]\neq1, 
\hspace{3mm}\bigl[[S_a],[S_b]\bigl]\neq0,
\end{equation}
and

\begin{equation}
(\gamma)=[\xi]\neq1.
\end{equation}
We thus see that a round trip along a trefoil knot $\gamma$ can cause 
some physical effects even if $\gamma$ surrounds an area on which 
$F_{\mu\nu}(x)$ vanishes. To be more specific, we consider the case 
that the gauge group is $SU(2)$ and that the $T^a$ in (1$\cdot$4) 
belongs to the fundamental representation. Then we can assume

\begin{equation}
[S_a]=e^{\mbox{\it i{\mathversion{bold}$A\cdot\sigma$}}},\hspace{3mm}
[S_b]=e^{\mbox{\it i{\mathversion{bold}$B\cdot\sigma$}}}
\end{equation}
where {\mathversion{bold}$\sigma$}$=(\sigma_1,\sigma_2,\sigma_3)$ with 
$\sigma_1,\sigma_2,\sigma_3$ being Pauli matrices, and 
\mbox{\bf\itshape A} and \mbox{\bf\itshape B} are three dimensional 
real vectors. Defining  \mbox{\bf\itshape K} by

\begin{equation}
[\xi]=e^{\mbox{\it i{\mathversion{bold}$K\cdot\sigma$}}},
\end{equation}
the formula for the Wilson loop 
$W(\gamma)\equiv\mbox{\rm tr}(\gamma)$ is given by 

\begin{equation}
W(\gamma)=2\cos K=2\bigl[1-2(\sin\varphi\sin\alpha\sin\beta)^2\bigl],
\end{equation}

\begin{equation}
\alpha=|\mbox{\bf\itshape A}|,\hspace{3mm}\beta=|\mbox{\bf\itshape B}|,
\hspace{3mm}\cos\varphi
=\frac{\mbox{\bf\itshape A}\cdot\mbox{\bf\itshape B}}{\alpha\beta},
\hspace{3mm}K=|\mbox{\bf\itshape K}|.
\end{equation}
We have thus explicitly seen that the loop variable $(\gamma)$ and 
the Wilson loop $W(\gamma)$ are nontrivial if $\alpha$, $\beta$ as 
well as $\varphi$ are not equal to an integer multiple of $\pi$.

\section{NAST for a Hopf link}

We next investigate the NAST in the case that the loop $\gamma$ consists 
of some connected components. The simplest case is a Hopf link which 
consists of two connected components $\gamma_1$ and ${\gamma_2}'$ 
as is shown in Fig.7(a) :

\begin{equation}
\gamma={\gamma_2}'\circ\gamma_1.
\end{equation}

\begin{figure}
    \epsfxsize=14cm
    \centerline{\epsfbox{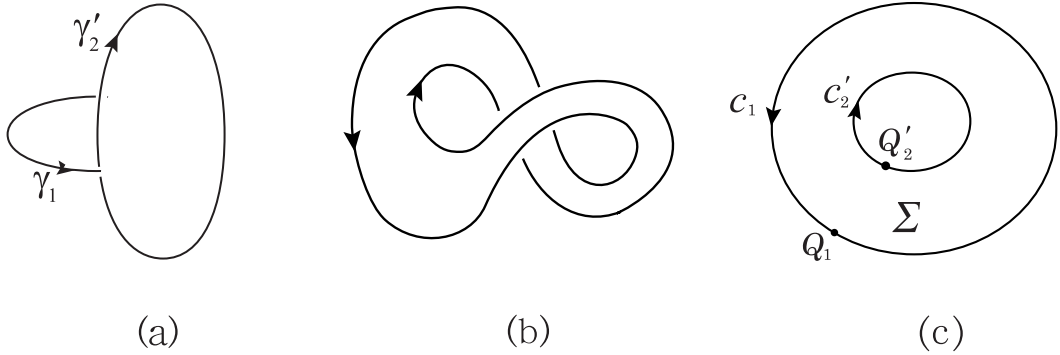}}
\caption{(a)A Hopf link, (b)A Seifert surfaces of a 
            loop ambient isotopic to a Hopf link,
          (c)The parameter space for a Seifert surface of a Hopf link.}
\label{Fig.7}
\end{figure}

One of the Seifert surfaces of $\gamma$ is given in Fig.7(b), which is 
homeomorphic to the doubly connected surface $\Sigma$ of Fig.7(c) satisfying
$\partial\Sigma={c_2}'\circ c_1$. We are allowed to assume 

\begin{equation}
\gamma_1=x(c_1),\hspace{3mm}{\gamma_2}'=x({c_2}'),
\end{equation}
where $x$ is the mapping from $\Sigma$ to the spacetime. The simply 
connected surfaces $\sigma_1$ and ${\sigma_2}'$ are
surrounded by $c_1$ and $\overline{{c_2}'}$,
respectively. They are related to $\Sigma$ by 

\begin{equation}
\Sigma=\sigma_1-{\sigma_2}'.
\end{equation}
If we set $S_1=x(\sigma_1)$, ${S_2}'=x({\sigma_2}')$ and $S=x(\Sigma)$,
we have 

\begin{equation}
S=S_1-{S_2}'.
\end{equation}
Supposing that the loop $c_1({c_2}')$ starts and ends at the point 
$Q_1({Q_2}')$ and denoting a path from $Q_1$ to ${Q_2}'$ by $d$,
we define $D$ by $D=x(d)$. Now the NAST of \S1 yields the following 
relations : 

\begin{equation}
\left.\begin{array}{l}
(\gamma_1)=[S_1],\\
\\
(\overline{{\gamma_2}'})=[{S_2}'],\\
\\
(D)^{-1}({\gamma_2}')(D)(\gamma_1)=[S].\\
\end{array}\right.
\end{equation}
We see that the NAST for the loop variable 

\begin{equation}
(\gamma)=({\gamma_2}')(\gamma_1)
\end{equation}
is given by

\begin{equation}
(\gamma)=\{S_1;{S_2}'\},
\end{equation}
where $\{S_1;{S_2}'\}$ is defined by 

\begin{equation}
\{S_1;{S_2}'\}=[{S_2}']^{-1}[S_1].
\end{equation}
We see that the simple result $(\gamma)=[S]$, (1$\cdot$10), for a trivial 
loop is violated also in this example.

If we consider the case that the parameter space $\Sigma$ is $\mu$-ply 
connected as is shown in Fig.8, we are led to 

\begin{equation}
(\gamma)=\{S_1;{S_2}',{S_3}',\cdots{S_\mu}'\}
\end{equation}
where $(\gamma)$ and $\{S_1;{S_2}',{S_3}',\cdots{S_\mu}'\}$ are defined by 

\begin{equation}
(\gamma)=({\gamma_\mu}')\cdots({\gamma_3}')({\gamma_2}')(\gamma_1),
\end{equation}

\begin{equation}
\{S_1;{S_2}',{S_3}',\cdots{S_\mu}'\}=[{S_\mu}']^{-1}\cdots[{S_3}']^{-1}
[{S_2}']^{-1}[S_1].
\end{equation}

\begin{figure}
    \epsfxsize=8cm
    \centerline{\epsfbox{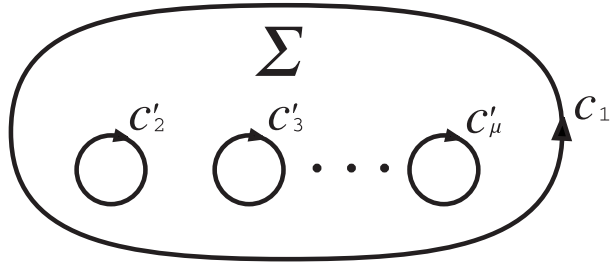}}
\caption{The parameter space of a link with $\mu$ connected components.}
\label{Fig.8}
\end{figure}

\section{NAST for general links}

In \S2 and \S3, we have considered the case of the simplest but nontrivial
examples of knots and links. In this section we shall obtain the NAST 
for a loop variable associated with a general link.

\subsection{Preliminaries\cite{rf:16,rf:17}}

Let us consider a compact orientable surface $F$ and a link $L$ in
$\mbox{\bf\itshape R}^3$. We say that $F$ is a Seifert surface of $L$ 
if the boundary of $F$ is equal to $L$ : $\partial F=L$.
When a Seifert surface consists of some connected components, 
we can make a connected Seifert surface by the procedure of the connected 
sum which does not violate the relation $\partial F=L$. 
So, if necessary, we can assume that the Seifert surface is connected.
The following theorem was discovered more than sixty years ago .\\
{\bf Theorem A}. Any oriented link has a Seifert surface.
 
If $F$ is a Seifert surface of a link $L$, the surface $F'$ obtained by 
the following procedure is also a Seifert surface of $L$ :

\begin{equation}
F'=(F-E_0-E_1)\cup h(S^1\times[0,1]),
\end{equation}
where $E_0$ and $E_1$ are two disks inside $F$ satisfying 
$E_0\cap E_1=\phi$ and $h(S^1\times[0,1])$ is a handle 
to be attached to the surface $F-E_0-E_1$ along 
$\partial E_0=h(S^1\times\{0\})$ and $\partial E_1=h(S^1\times\{1\})$. 
If the handle $h(S^1\times[0,1])$
is attached on one side of $F-E_0-E_1$, the orientation of $F'$ 
is naturally induced from that of $F$. For the above $F$ and $F'$, 
we say that $F'$ is obtained from $F$ by a 1-surgery. 
Conversely we say that $F$ is obtained
from $F$ by a 0-surgery. The genus of a connected orientable surface $F$ 
is given by 

\begin{equation}
g(F)=\frac{1}{2}\bigl(2-\chi(F)-\mu\bigl),
\end{equation}
where $\mu$ is the number of the boundary components of $F$ and $\chi(F)$ 
is the Euler characteristic of $F$. We easily see 

\begin{equation}
g(F')=g(F)+1. 
\end{equation}
We then understand that, for a prescribed link, 
there are many Seifert surfaces with various values of genus. 
Among them a surface with the smallest genus is called the minimum 
Seifert surface. If two surfaces $F$ and $F'$ are obtained by some steps of 
0-and/or 1-surgeries from each other, they are said to be stably equivalent
to each other.
It can be seen that a 1-surgery of the surface $F$ is equivalent to attaching 
the closed surface $\overline{E_0}\cup h(S^1\times[0,1])\cup\overline{E_1}$
to $F$, where $\overline{E}$ is equal to $E$ with the orientation reversed.
We have 

\begin{equation}
[\overline{E_0}\cup h(S^1\times[0,1])\cup\overline{E_1}]=1
\end{equation}
since the surface $\overline{E_0}\cup h(S^1\times[0,1])\cup\overline{E_1}$
is homeomorphic to a sphere and we know [sphere]=1.
Furthermore there exists the following remarkable 
theorem. \\
{\bf Theorem B}. Any two connected Seifert surfaces of an oriented link
are stably equivalent to each other.

The genus $g(L)$ of a link $L$ is defined by $g(L)=g(F)$, where $g(F)$ is 
the genus of the minimum Seifert surface $F$ of $L$. As was stated in the 
above, we can assume that $F$ is connected. We here cite the classification
theorem of surfaces.\\
{\bf Theorem C}. Any connected orientable surface with boundaries is 
homeomorphic to one of $T(g,\mu)$, $g=0,1,2,\cdots,$ $\mu=1,2,3,\cdots,$
where $T(g,\mu)$ is given in Fig.9.

\begin{figure}
    \epsfxsize=14cm
    \centerline{\epsfbox{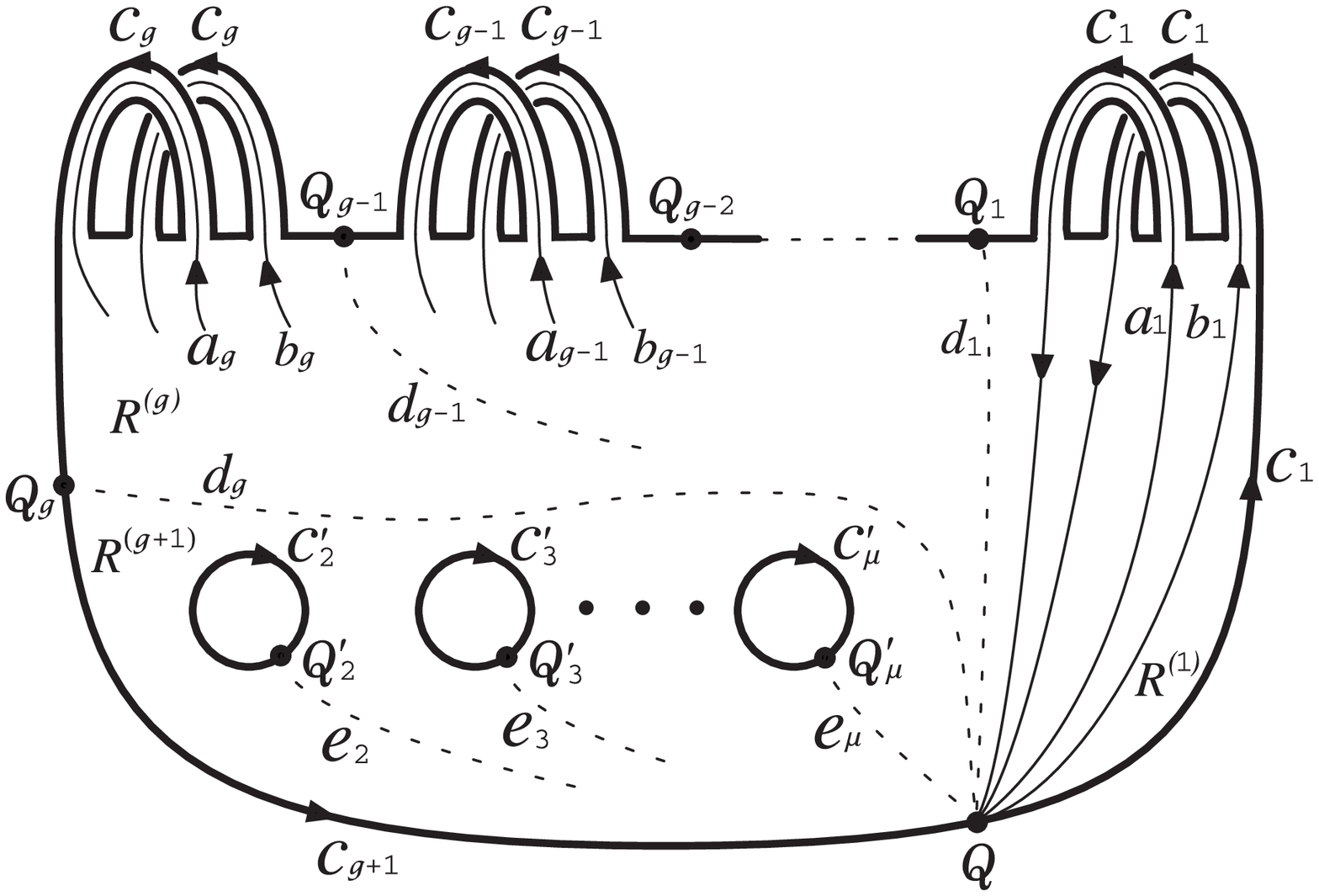}}
\caption{$T(g,\mu)$.}
\label{Fig.9}
\end{figure}

We note that $T(0,1)$ is a disk and that $T(1,1)$ and $T(0,\mu)$, 
$\mu\geq2$, are given by Fig.3 and Fig.8, respectively.

\subsection{Derivation of NAST for a general link}

Suppose that a link $\gamma$ with the genus $g$ consists of 
$\mu$ connected components. 
We adopt one of the minimum Seifert surface of $\gamma$ and 
denote it by $S$. Then the surface $S$ can be expressed as 

\begin{equation}
S=\{x(s,t)|(s,t)\in\Sigma\},\hspace{3mm}\Sigma=T(g,\mu),
\end{equation}
where $x$ is, as in the previous sections,
a continuous mapping from the parameter space $\Sigma$ to the
spacetime. The curves $a_i$, $b_i$, $i=1,2,\cdots,g,$ and $e_\alpha$,
$\alpha=2,3,\cdots,\mu$, in Fig.9 are helpful for the discussion of
the loop variable $(\gamma)$. We assume that the link $\gamma$ 
is ordered as

\begin{equation} 
\left.\begin{array}{l}
\gamma=x(\partial\Sigma)\\
\\
\hspace{3mm}={\gamma_\mu}'\circ{\gamma_{\mu-1}}'\circ\cdots\circ{\gamma_3}'
\circ{\gamma_2}'\circ\gamma_{g+1}\circ\gamma_g\circ\cdots\circ\gamma_2
\circ\gamma_1\\
\end{array}\right.
\end{equation}
with $\gamma_i$ and ${\gamma_\alpha}'$ given by 

\begin{equation} 
\left.\begin{array}{l}
\gamma_i=x(c_i),\hspace{3mm}i=1,2,\cdots,g+1,\\
\\
{\gamma_\alpha}'=x({c_\alpha}'),\hspace{3mm}\alpha=2,3,\cdots,\mu,\\
\end{array}\right.
\end{equation}
where the curves $c_i$ and ${c_\alpha}'$ constitute the boundary of $\Sigma$ :
$\partial\Sigma={c_\mu}'\circ{c_{\mu-1}}'\circ\cdots\circ{c_3}'\circ
{c_2}'\circ c_{g+1}\circ c_g\circ\cdots\circ c_2\circ c_1$. 
With the help of the auxiliary curves 
$d_i$ $(i=1,2,\cdots,g)$ starting at $Q_i$ and ending at $Q$,
the surface $\Sigma$ is
devided into $g+1$ areas. The areas surrounded by $d_1\circ c_1$, 
$d_i\circ c_i\circ\overline{d}_{i-1}$ $(i=2,3,\cdots,g)$ and 
${c_\mu}'\circ\cdots\circ{c_3}'\circ{c_2}'\circ c_{g+1}\circ\overline{d_g}$
are denoted by $R_1$, $R_i$ $(i=2,3,\cdots,g)$ and $R_{g+1}$, respectively.
We then have 

\begin{equation} 
\left.\begin{array}{l}
\partial R_1=d_1\circ c_1,\\
\\
\partial R_i=d_i\circ c_i\circ\overline{d_{i-1}},
\hspace{3mm}i=2,3,\cdots,g,\\
\\
\partial R_{g+1}
={c_\mu}'\circ\cdots\circ{c_3}'\circ{c_2}'\circ c_{g+1}\circ\overline{d_g}.\\
\end{array}\right.
\end{equation}
We see that the method of \S2 (\S3) can be applied to $x(R_i)$ and 
$x(\partial R_i)$, $i=1,2,\cdots,g$, $(i=g+1)$. Defining $S^{(i)}$ by

\begin{equation} 
S^{(i)}=x(R_i),\hspace{3mm}i=1,2,\cdots,g,g+1,
\end{equation}
and dividing $S^{(i)}$, $i=1,2,\cdots,g$, into four areas,

\begin{equation}
S^{(i)}=\bigcup_{k=1}^4S^{(i)}_k,
\end{equation}
as in \S2, we have 

\begin{equation}
(D_1)(\gamma_1)=[\hspace{-0.7mm}[S^{(1)}]\hspace{-0.7mm}],
\end{equation}

\begin{equation}
(D_i)(\gamma_i)(D_{i-1})^{-1}=[\hspace{-0.7mm}[S^{(i)}]\hspace{-0.7mm}],
\hspace{3mm}i=2,3,\cdots,g,
\end{equation}
where $D_i$ and $[\hspace{-0.7mm}[S^{(i)}]\hspace{-0.7mm}]$ are defined by

\begin{equation}
D_i=x(d_i),
\end{equation}

\begin{equation}
[\hspace{-0.7mm}[S^{(i)}]\hspace{-0.7mm}]=[S_{a_i}][S_4^{(i)}][S_{b_i}]^{-1}
[S_3^{(i)}][S_{a_i}]^{-1}[S_2^{(i)}][S_{b_i}][S_1^{(i)}].
\end{equation}
We also have 

\begin{equation}
({\gamma_\mu}')({\gamma_{\mu-1}}')\cdots({\gamma_2}')(\gamma_{g+1})
(D_g)^{-1}=\{S^{g+1};{S_2}',{S_3}',\cdots,{S_\mu}'\},
\end{equation}
where the r.h.s. is defined in a similar manner to (3$\cdot$11).
From (4$\cdot$6), (4$\cdot$11), (4$\cdot$12) (4$\cdot$15), we finally obtain

\begin{equation}
(\gamma)=\{S^{g+1};{S_2}',{S_3}',\cdots,{S_\mu}'\}
[\hspace{-0.7mm}[S^{(g)}]\hspace{-0.7mm}]
[\hspace{-0.7mm}[S^{(g-1)}]\hspace{-0.7mm}]\cdots
[\hspace{-0.7mm}[S^{(2)}]\hspace{-0.7mm}]
[\hspace{-0.7mm}[S^{(1)}]\hspace{-0.7mm}],
\end{equation}
which is the NAST for a general link with the $\gamma$ ordered by (4$\cdot$6).
If the ordering of $\gamma$ is different from that of (4$\cdot$6), 
the r.h.s. of (4$\cdot$16) must be replaced by an expression in which the 
ordering of $[\hspace{-0.7mm}[*]\hspace{-0.7mm}]$'s and $\{*\}$'s is changed.

\subsection{Independance of $(\gamma)$ on the choice of S}

We are left with the problem to show that the r.h.s. of (4$\cdot$16)
is independent of the choice of the Seifert surface $S$. With the help 
of the theorem B, it can be seen that the problem is reduced to show
the equality

\begin{equation}
[x(\Sigma)]=[x(\Sigma_1)].
\end{equation}
Here the parameter space $\Sigma$ and $\Sigma_1$ are those shown in Fig.10, 
$\Sigma_1$ being  obtained from $\Sigma$ through a 1-surgery. 

\begin{figure}
    \epsfxsize=8cm
    \centerline{\epsfbox{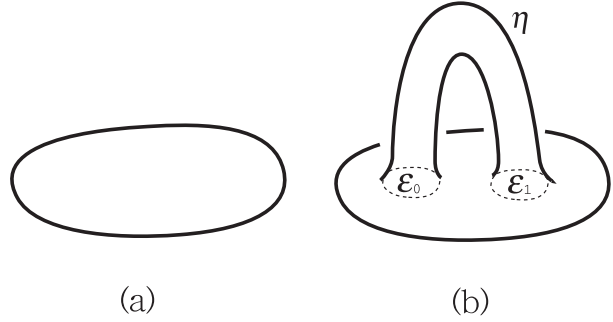}}
\caption{(a)$\Sigma$, (b)$\Sigma_1=(\Sigma$ with a handle attached).}
\label{Fig.10}
\end{figure}

The r.h.s. of (4$\cdot$17) is somewhat symbolical since the surface 
$\Sigma_1$ is not simply connected. Its meaning becomes unambiguous 
only after an indication of the ordering is given. Just as in the case of
Eq.(4$\cdot$1), the surface $\Sigma_1$ can be regarded to consist of two 
disks $\overline{\varepsilon_0}$ and $\overline{\varepsilon_1}$, 
a handle $\eta$ and the surface $\Sigma$ : 
$\Sigma_1=(\overline{\varepsilon_0}\cup\eta\cup\overline{\varepsilon_1})
\cup\Sigma$. The ordering for $[x(\Sigma_1)]$ can be prescribed by 

\begin{equation}
[x(\Sigma_1)]
=[x(\overline{\varepsilon_0}\cup\eta\cup\overline{\varepsilon_1})]
[x(\Sigma)].
\end{equation}
Since the first factor of the r.h.s. of (4$\cdot$18) is equal to 1
as was explained below (4$\cdot$4), we are led to (4$\cdot$17).

\section{Summary}

In this paper we have sought the NAST for loop variables associated with 
nontrivial loops. It turned out that the case of the trefoil knot (Fig.1) is 
of fundamental importance and constitutes the building block of the 
general case. The NAST for this case is given by (2$\cdot$10), 
where the quantities $[S_a]$and $[S_b]$ appear in addition to 
$[S_i]$, $i=1,2,3,4$. Another expression of the NAST is given by 
(2$\cdot$13), in which the factor $[\xi]$ defined by (2$\cdot$14) appears. 
Thanks to the deformation invariance of $[S]$, (1$\cdot$14),    
and the theorem B of \S4, 
the result does not depend on the choice of a Seifert surface 
of the trefoil knot.    
The structure of the NAST for the case of the figure eight
knot (Fig.5) is the same as that of the trefoil knot since the parameter
space for these two cases can be chosen homeomorphic
to each other. We have seen, in sharp contrast to the Abelian case, that 
the loop variable $(\gamma)$ can be different from unity even if the field 
strength vanishes evrywhere on the surface surrounded by the loop $\gamma$.
We expect that the above fact might cause some interesting physical effects.

The NAST for a generic link of genus $g$ consisting of $\mu$ connected
components was simply expressed with the help of the quantities 
$[\hspace{-0.7mm}[*]\hspace{-0.7mm}]$ and $\{*\}$ defined by (4$\cdot$14)
and (3$\cdot$11), respectively.

\section*{Acknowledgements}

The authors are grateful to S.Hamamoto, T.Kurimoto and S.Matsubara
for their kind interest in this work.

\end{document}